\documentclass[aps,prd,twocolumn,nofootinbib,superscriptaddress]{revtex4-1}

\usepackage[utf8]{inputenc}
\usepackage{orcidlink}
\usepackage{color}
\usepackage{graphicx}
\usepackage{amsmath}
\usepackage{amssymb}
\usepackage{bm}
\usepackage{acronym}
\usepackage{ifthen}
\usepackage{blindtext}
\usepackage[normalem]{ulem}
\usepackage{hyperref}
\usepackage{etoolbox}
\usepackage{fancyhdr}
\usepackage{xspace}
\usepackage{textcomp}
\usepackage{multirow}
\usepackage[caption=false]{subfig}
\usepackage{tabularx}
\usepackage{dcolumn}
\usepackage[mathlines]{lineno}
\graphicspath{{./figures/}}

\newtoggle{fullauthorlist}
\toggletrue{fullauthorlist}
\newtoggle{endauthorlist}
\toggletrue{endauthorlist}

\begin{document}


\title{Search for continuous gravitational wave emission from the Milky Way center in O3 LIGO--Virgo data}
\date{\today}

 \iftoggle{endauthorlist}{
  %
  %
  \let\mymaketitle\maketitle
  \let\myauthor\author
  \let\myaffiliation\affiliation
  \author{The LIGO Scientific Collaboration}
  \author{The Virgo Collaboration}
  \author{The KAGRA Collaboration}
  \email{Full author list given at the end of the article.}
\noaffiliation
}{
\iftoggle{fullauthorlist}{


\author{R.~Abbott}
\affiliation{LIGO Laboratory, California Institute of Technology, Pasadena, CA 91125, USA}
\author{H.~Abe}
\affiliation{Graduate School of Science, Tokyo Institute of Technology, Meguro-ku, Tokyo 152-8551, Japan  }
\author{F.~Acernese}
\affiliation{Dipartimento di Farmacia, Universit\`a di Salerno, I-84084 Fisciano, Salerno, Italy  }
\affiliation{INFN, Sezione di Napoli, Complesso Universitario di Monte S. Angelo, I-80126 Napoli, Italy  }
\author{K.~Ackley\,\orcidlink{0000-0002-8648-0767}}
\affiliation{OzGrav, School of Physics \& Astronomy, Monash University, Clayton 3800, Victoria, Australia}
\author{N.~Adhikari\,\orcidlink{0000-0002-4559-8427}}
\affiliation{University of Wisconsin-Milwaukee, Milwaukee, WI 53201, USA}
\author{R.~X.~Adhikari\,\orcidlink{0000-0002-5731-5076}}
\affiliation{LIGO Laboratory, California Institute of Technology, Pasadena, CA 91125, USA}
\author{V.~K.~Adkins}
\affiliation{Louisiana State University, Baton Rouge, LA 70803, USA}
\author{V.~B.~Adya}
\affiliation{OzGrav, Australian National University, Canberra, Australian Capital Territory 0200, Australia}
\author{C.~Affeldt}
\affiliation{Max Planck Institute for Gravitational Physics (Albert Einstein Institute), D-30167 Hannover, Germany}
\affiliation{Leibniz Universit\"at Hannover, D-30167 Hannover, Germany}
\author{D.~Agarwal}
\affiliation{Inter-University Centre for Astronomy and Astrophysics, Pune 411007, India}
% ...

\collaboration{The LIGO Scientific Collaboration, the Virgo Collaboration, and the KAGRA Collaboration}



}
{
  \author{The LIGO Scientific Collaboration}
  \affiliation{LSC}
  \author{The Virgo Collaboration}
  \affiliation{Virgo}
  \author{The KAGRA Collaboration}
  \affiliation{KAGRA}
  
}
}  
\begin{abstract}
We present a directed search for continuous gravitational wave (CW) signals emitted by spinning neutron stars located in the inner parsecs of the Galactic Center (GC). Compelling evidence for the presence of a numerous population of neutron stars has been reported in the literature, turning this region into a very interesting place to look for CWs.
In this search, data from the full O3 LIGO--Virgo run in the detector frequency band $[10,2000]\rm~Hz$ have been used.  
No significant detection was found and 95$\%$ confidence level upper limits on the signal strain amplitude were computed, over the full search band, with the deepest limit of about $7.6\times 10^{-26}$ at $\simeq 142\rm~Hz$. These results are significantly more constraining than those reported in previous searches. We use these limits to put constraints on the fiducial neutron star ellipticity and r-mode amplitude. These limits can be also translated into constraints in the black hole mass -- boson mass plane for a hypothetical population of boson clouds around spinning black holes located in the GC.

\end{abstract}

\maketitle

\section{Introduction}
The Milky Way's center is one of the most interesting sky regions and is well suited to investigations with multiple messengers, ranging from electromagnetic radiation and cosmic rays to gravitational waves (GWs). 
A significant population of up to hundreds or even thousands of neutron stars is expected to exist in this region, based on the observation of high-mass progenitor stars, and of several supernova remnants \cite{Rajwade2017,Kim2018}.   
Moreover, an extended gamma-ray emission from the central region of the Galaxy has been detected by the Fermi Large Area Telescope \cite{FermiGCE2016,FermiGCE2021} and H.E.S.S. \cite{HESS2016}. The true nature of this emission is still under debate, with two main competing explanations: annihilation of dark matter in the form of Weakly Interacting Massive Particles, with masses of the order of a few tens of GeV \cite{CuocoDM2017,CuiDM2017,AlbertDM2017,AckermannDM2017,DiMauroDM2021}, or an unresolved population of millisecond pulsars \cite{Bartels_2016, Lee_2016,Calore2016,Gregoire2013,Fermi2017characterizing,Hooper2018,Buschmann2020}. This diffuse emission may also actually be due to a combined contribution from both a population of millisecond pulsars and heavy dark matter \cite{Lacroix2016}. 

The possibility that the GC region hosts a large population of neutron stars calls for the search for CWs that neutron stars would emit if their shape deviated from axial symmetry. Although previous searches have not reported any detection (see the reviews  \cite{Piccinni2022status,Riles_2017,Sieniawska2019} and references therein plus the last results in \cite{LVKtargetedO32020,LVKbinaryallskyO3a2021,LVKtargetedO3J053769102021,LVKrmodeO32021constraints,SNRO3,LVKallskyO32021,LVKbinariesO32021}), improvements in detectors and data analysis pipelines \cite{Tenorioreview2021} can increase the search sensitivity to a level at which detections can take place.

In this work we present the results of a search, using the latest data from the third observing run (O3) of the Advanced Virgo \cite{AdVirgo} and Advanced LIGO \cite{AdLIGO} detectors, for CWs emitted by non-axisymmetric rotating neutron stars located in the GC region. This work, based on the Band-Sampled-Data (BSD) directed search pipeline\cite{BSD,DirectedBSD}, improves over a previous search in O2 data \cite{DirectedBSD}.

O3 data have been already used to make a lower sensitivity search for stochastic GW emission from the same region \cite{LVCstochanis2021}. 

Potentially detectable CW emission is expected from galactic, fast-spinning neutron stars with a certain degree of asymmetry in their mass distribution \cite{Glampedakis_2018,Sieniawska2019}. 
As the star spins, it releases energy in the form of GWs, which are almost monochromatic and characterized by an emitted frequency proportional to the star’s spin frequency. 
The frequency received at a detector evolves in time due to two main contributions. One comes from the intrinsic frequency decrease (spin-down) caused by energy loss of the star. The other is due to the Doppler modulation of the received signal, caused by the motion of the detector with respect to the source. Other smaller effects are also considered, namely the Einstein and the Shapiro delays.

The asymmetry in the neutron star's shape can be due to different reasons.  They include the presence of a residual crustal deformation (e.g., after a fast cooling of the neutron star crust causing its breaking), the presence of a strong internal magnetic field not aligned with the star's rotation axis, or the presence of magnetic or thermal ``mountains”. The maximum ellipticity (i.e., degree of deformation) a neutron star can sustain depends on both its equation of state (EOS) and the breaking strain of the crust \cite{Johnson-McDaniel2013}. On the other hand, the typical degree of asymmetry is difficult to estimate. The detection of a CW signal would help to shed light on the internal structure of neutron stars, hence on the EOS, given the relation of the GW amplitude to the star’s moment of inertia and the star’s ellipticity. GWs are then unique probes of the fundamental interactions happening inside a neutron star, making them ideal laboratories to test fundamental physics and high-energy astrophysics in the strong-gravity regime \cite{Lattimer_2004}. A different emission channel for CW radiation is given by the Rossby (r-)modes oscillations in rotating stars \cite{Andersson1998,Bildsten1998,Friedman1998}.

The emission of CWs is also expected from other sources, not involving neutron stars. For instance, it has been predicted \cite{Arvanitaki2015,Baryakhtar2021,BritoSuperradiance2020,Brito:2017zvb} that if ultra-light boson particles (a dark matter candidate) exist in nature, they may spontaneously form macroscopic ``clouds'' around spinning black holes through a \textit{superradiance} process (provided that the black hole initial spin is high enough). Once formed such clouds dissipate through the emission of a CW signal, with a frequency mainly depending on the boson mass. 
In this work we also use the CW search results to constrain a possible boson cloud population in the GC.\\

The paper is organized as follows. In Sec. \ref{sec:data}, the data used for the analysis are described. Sec. \ref{sec:search} is devoted to discussing the  explored parameter space and the main search method. Post-processing is presented in Sec. \ref{sec:postprocessing}, including all the vetoes used to discard instrumental artefacts. In Sec. \ref{sec:result}, we present the search results and their astrophysical implications, for spinning neutron stars and for boson clouds around spinning black holes. In Sec. \ref{sec:conc}, conclusions are drawn. Some details on the upper limits computation are given in Appendix \ref{sec:appen}.

\section{The data}
\label{sec:data}
For this search we have used the full O3 data of the two Advanced LIGO detectors in Hanford (H) and Livingston (L) in the United States \cite{AdLIGO} and of Advanced Virgo (V) in Cascina, Italy \cite{AdVirgo}.  The O3 observing run started on April 1st, 2019 at 15:00 UTC and ended on March 27th, 2020 at 17:00 UTC.   During data taking there was a one-month break, from October 1st, 2019 to November 1st, 2019.  The duty factors for O3 were $76\%$, $71\%$, $76\%$ for L, H, V respectively. The sensitivities of the three detectors are comparable at lower frequencies, while Virgo sensitivity above $\sim 100~\rm Hz$ is smaller than the other two LIGO detectors.
The last version of the high-latency calibrated data (C01 frames)~\cite{SunO3acalibration2020} has been used for H and L,  while for Virgo the ``online" calibration version has been used. Moreover, only
science segments, i.e. time intervals when the detector is operating in a nominal state and the noise level is considered as acceptable, have been selected.
The maximum calibration amplitude uncertainties for LIGO are $7\%$ during the first half of O3 (O3a)~\cite{SunO3acalibration2020}, and  $11\%$ during the second half of O3 (O3b)~\cite{SunO3bcalibration2021}. For Virgo the calibration uncertainty is $5\%$ in amplitude for the full frequency band except for regions around 50 Hz where larger uncertainties appear~\cite{AcerneseO3calibration2021}. 
A gating procedure has been applied to LIGO data as described in~\cite{T2000384} to remove larger transient artifacts.
Furthermore, for all the detectors, short-duration noise transients have been removed during the construction of the Short Fourier Transform Database (SFDB)~\cite{FFTpeakmaps}. The search is based on the BSD framework~\cite{BSD},  which works with time series sampled at 10 Hz and spanning a 10 Hz frequency band, computed from the SFDB. Indeed in the BSD framework, the parameter space investigated, as well as the choice of the grid steps and the coherence time used, change every 10 Hz. Given the limited computational power available, the parameter space investigated has been chosen satisfying some constraints described in Sec. \ref{ssec:parspace}.
Spectral noise artifacts, known as lines, are also present in detector data. Lists of narrow lines with identified instrumental origin are given in \cite{linelist01,virgo3lines}.  These lists will be used in the post-processing stage, to veto candidates near instrumental lines (see Sec. \ref{ssec:linesveto}).

\section{The search}
\label{sec:search}
The search is conducted with a semi-coherent method \cite{DirectedBSD} in which the data are divided into segments of given duration $T_{\rm coh}$. These segments are properly processed, as will be clarified in subsection \ref{ssec:method}, and then incoherently combined, i.e. not taking into account the signal phase. In this way we can explore a large parameter space at a fixed computing cost, with only a relatively small sensitivity loss with respect to an optimal, fully coherent procedure when applied in an all-sky search (see e.g. Sec. XIID in \cite{FrequencyHoughmethod}). In this search we do not explicitly search for CW signals from neutron stars with a binary companion. However, as discussed in \cite{Singh2019}, the results presented in this work are valid to some extent also for accreting binaries.
 
\subsection{Parameter space}
\label{ssec:parspace}   
\begin{figure}
\includegraphics[scale=0.45]{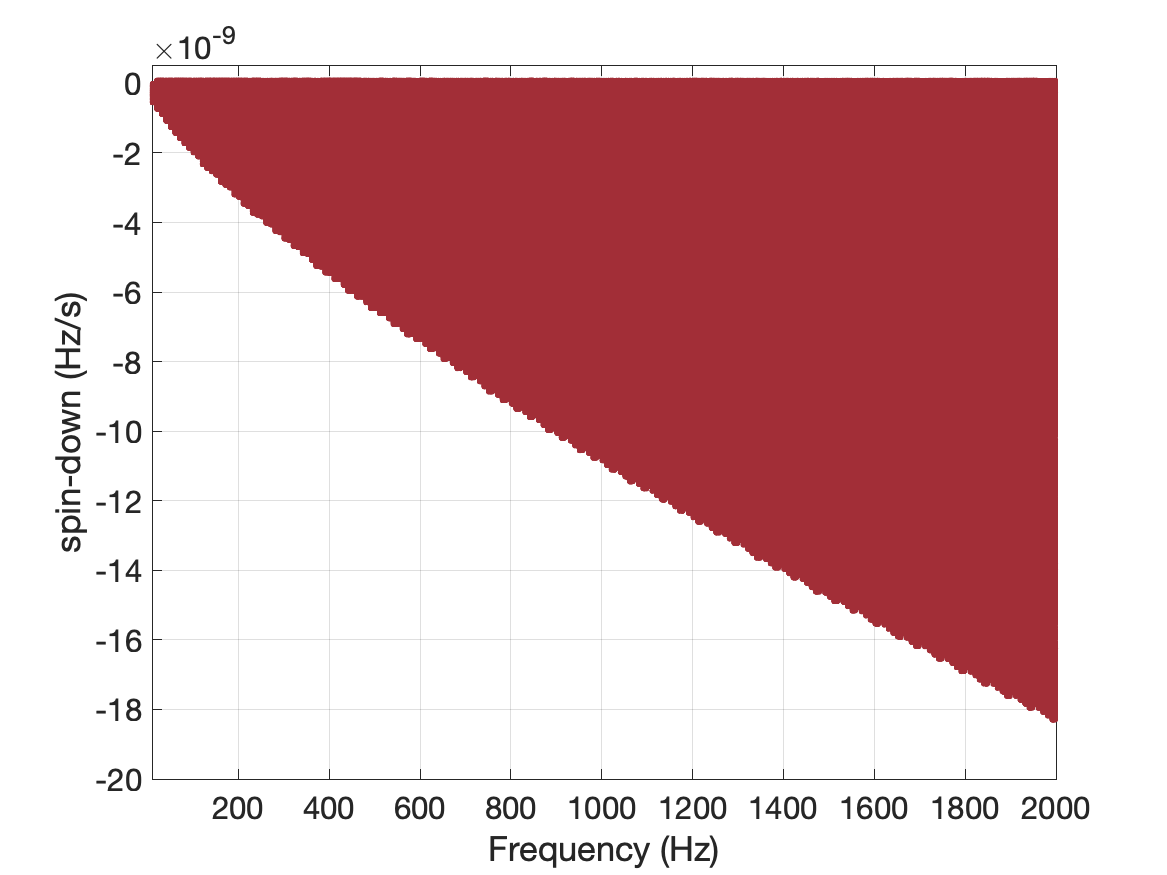}
\caption{Parameter space investigated in this search (see Sec. \ref{ssec:parspace}). The sky position is that of Sgr A*, while frequency and spin-down span the ranges given on the axes.}
\label{fig:parspace}
\end{figure}

We look for persistent signals from sources emitting in the detector frequency band $[10,2000]\rm~Hz$ and located in the GC, with a maximum spin-down range of $[-1.8 \times 10^{-8},~10^{-10}] \,\rm\,Hz/s$.
The actual minimum value of the spin-down is a function of the search frequency, as we will clarify below; see also Fig.  \ref{fig:parspace}.

The parameter-space volume is discretized in several cells. The key quantity governing the discretization is the segment duration $T_{\rm coh}$.
Its value is chosen as a function of the frequency to keep the signal frequency at the detector, which varies due to the spin-down and Doppler effect, within one frequency bin, given by $\delta f=1/T_{\rm coh}$. In each 10-Hz band, the smallest $T_{\rm coh}$ is adopted for the whole band.  
The corresponding number of frequency bins is $N_f=10\rm~Hz\times T_{\rm coh}$. 
The spin-down bin size is given by
\begin{equation}
    \delta \dot{f} = \frac{1}{2T_{\rm obs} T_{\rm coh}}
\end{equation}
where $T_{\rm obs}$ is the total observing time. The bin width for the second order spin-down is
\begin{equation}
    \delta \ddot{f} = \frac{1}{4T^2_{\rm obs} T_{\rm coh}}.
\end{equation}
To minimize the computational load of the analysis, we consider a single value for the second order spin-down, i.e. $N_{\ddot{f}}=1$. The number of second order spin-down bins depends on the parameter space investigated and the search setup, $N_{\ddot{f}}=2N_{f}\left(T_{\rm obs}|\dot{f}|/f\right)^{2}$, which always satisfies the condition $N_{\ddot{f}}\leq1$ for this search.
For balancing the computational cost among the available resources, and taking into account the available CPU memory, a different range of spin-down values is computed for every 10-Hz band, resulting in a smaller minimum value at low frequencies. The number of spin-down/spin-up bins $N_{\dot{f}}$ is then given by the $\dot{f}$ range covered in each 10-Hz band divided by the corresponding $\delta{\dot{f}}$.
The minimum spin-down goes from $-5.6\times 10^{-10}$ Hz/s in the band 10--20 Hz to $-1.8\times 10^{-8}$ Hz/s in the band 1990--2000 Hz band. This choice corresponds to a value of a characteristic spin-down age\footnote{This is a rough measure of the star's age, which assumes the initial spin frequency is much higher than the current one.} $\tau=f/|\dot{f}|$ that goes from $\simeq 570$ years at 10 Hz to $\simeq 3520$ years at 2000 Hz.  
The small range of positive $\dot{f}$, from zero to $10^{-10}\,\rm\,Hz/s$ in the full frequency band, allows us to take into account a possible spin-up, as expected for the signal emitted by boson clouds around spinning black holes.
The frequency/spin-down ranges covered by the search are shown in Fig.
\ref{fig:parspace}. The estimated computing cost per detector is $\sim500$ core hours for jobs running on a Intel ES-2640V4 CPU. 

Concerning the sky, only one bin centered at the position of Sgr A* \cite{SgrAsky2020} with right ascension $\alpha_{\rm GC}$ = 17h 45m 40.04s, and declination $\delta_{\rm GC} = -29^{\circ} \,00'\, 28.1"$ is taken into account. With the values of $T_{\rm coh}$ we are using, one sky bin covers 30--300\, pc, depending on the frequency \cite{DirectedBSD}, which is wide enough to include the most interesting region around the GC. 

\subsection{Method}
\label{ssec:method}
For this work we use the hierarchical semi-coherent BSD directed search pipeline \cite{DirectedBSD} based on the FrequencyHough (FH) transform \cite{FrequencyHoughmethod,FrequencyHough}, recently used in the search for CW signals from supernova remnants in our Galaxy \cite{SNRO3} and previously used for a GC search in Advanced LIGO O2 data \cite{DirectedBSD}.

In this search, we first partially correct the time series, separately for each detector, removing the Doppler modulation in each 1-Hz frequency sub-band, using its central frequency as a reference (see \cite{DirectedBSD} for more details).  This is achieved by an \textit{heterodyne} correction,  i.e. multiplying the time series by a complex exponential factor $e^{i\Delta\phi}$, where $\Delta\phi$ is the signal phase variation associated to the Doppler effect and referring to the central frequency of each 1-Hz band within a 10-Hz band.

The second step of the search consists in the selection of the most significant peaks (collectively called ``peakmap'') in the time-frequency plane. In order to do this, first \textit{equalized} spectra~\cite{FFTpeakmaps} are evaluated. They are obtained as the ratio among the periodogram, given by the square modulus of the Fourier Transform of each data segment with duration $T_{\rm coh}$, and an average spectrum estimation. On the ratio, all the local maxima above a given threshold are selected. Each of these ``peaks'' is defined by a frequency and by the initial time of the corresponding data segment of length $T_{\rm coh}$. Each peakmap covers 10 Hz in frequency and one month in time.

The peakmap is then passed to the third step of the analysis, which consists in the FH transform. 
The FH transform maps each time-frequency peak to the intrinsic source frequency and spin-down $(f_0,\dot{f}_0)$ plane at a given reference time $t_0$. As described in \cite{FrequencyHoughmethod} each time-frequency peak in the peakmap becomes a line in the FH plane. Hence, each pixel of the FH map has an associated number count $n$, corresponding to the number of lines intersecting in that point.
The resolution of a single FH map is related to the coherence time $T_{\rm coh}$ used for the construction of the peakmaps.  The resolutions in frequency and spin-down are given by
\begin{eqnarray}
\delta f_{\rm FH} = \frac{1}{T_{\rm coh}K_{f}},
\\
\delta \dot{f}_{\rm FH}=\frac{1}{T_{\rm coh}T_{\rm obs}K_{\dot{f}}}.
\end{eqnarray}
The parameters $K_{f}$ and $K_{\dot{f}}$ are the over-resolution factors as described in \cite{FrequencyHoughmethod}, here chosen as $K_{f}=10$ and $K_{\dot{f}}=2$. 

For each 10-Hz band, this process is repeated for each month of data and the single FH maps are summed up into a final map. 
On this we select outliers, i.e. pairs $(f_0,\dot{f}_0)$ with high significance. More specifically, for each 0.01 Hz subband, and for each spin-down subband\footnote{To minimize the computational load of each analysis job, the full spin-down range has been divided into a few subbands, going from 9 for all the bands above 20 Hz to 16 jobs for the band 10--20 Hz.}, we take the two strongest outliers. In this way a maximum of $\sim 2000$ outliers are selected in each map, and for each spin-down subband. The total number of outliers in the search is $3410398$ for L,  3409029 for H and 3408485 for V. 
The strength of an outlier is expressed through the Critical Ratio (CR) defined as 
\begin{equation}
\rho_{\rm CR}(n)=\frac{n-\mu_{n}}{\sigma_{n}},
\end{equation}
where $\mu_{n}$ and $\sigma_{n}$ are the mean and the standard deviation of the FH number count $n$, respectively. 
The CR of an outlier corresponds to a significance, expressed e.g. in terms of p-values, compared to the expected CR distribution under the hypothesis that the data do not contain any signal and assuming a Gaussian noise distribution. 



Coincidences among pairs and triplets of outliers found in the data of the three detectors are computed. Coincidences among two outliers are defined on the base of a dimensionless distance defined as \cite{FrequencyHoughmethod}
\begin{equation}
\label{Eq:distance}
d=\sqrt{\left(\frac{\Delta f}{\delta f_{\rm FH}}\right)^2+\left(\frac{\Delta \dot{f}}{\delta \dot{f}_{\rm FH}}\right)^2},
\end{equation}
where $\Delta f$ and $\Delta \dot{f}$  are the differences between the outlier parameters. All the pairs of outliers from two different detectors (i.e. HL, HV, LV) with a distance smaller than $d_{\rm{thr}}=4$ pass to the post-processing stage \cite{SNRO3,DirectedBSD}. Moreover, mean parameters of coincident HL outliers are used to compute the distance from V outliers, using the same criterion as before, selecting in this way triple coincidences, which are subject to a similar post-processing. 

A total of 2142 triple HLV coincidences have been found, mostly from candidates with spin-up. Double coincidences are 37570 for HL, 36988 for HV and 37455 for LV. Given the lower sensitivity of the Virgo detector compared to Hanford and Livingston at higher frequencies, we only keep those double-coincident candidates below 100 Hz, reducing the number of HV and LV coincidences to 381 and 458 respectively. This choice is justified by the fact that a real signal showing a significant result in the Virgo detector above 100 Hz, should necessarily be evident also in the HL coincidence set. 

To screen out insignificant outliers, we select only the double coincidences with a $\rho_{\rm CR}(n) > \rho_{\rm CR,thr}$. 
The threshold is chosen considering the probability of picking, on average over each 10-Hz band, 500 false single-detector candidates over the total number of spin-down bins, under the assumption of Gaussian noise. 
The threshold used is $\rho_{\rm CR,thr}\simeq 4.78$ and it is the same for each dataset. This threshold is also very close to the mean $\rho_{\rm CR}$ plus one standard deviation of the CR distribution  across the single-detector candidates excluding those due to known instrumental lines. This step is described in Sec. \ref{ssec:linesveto}. 
From a computational point of view, using a higher threshold would have certainly reduced the number of potential candidates for further investigation. 
However, this would also increase the false-dismissal probability.
For the triple HLV coincidence set we focused on those candidates above $\rho_{\rm CR,thr}$, including those with a frequency below 100 Hz, independently of their CR.

\section{Post-processing}
\label{sec:postprocessing}
Before passing to the followup stage, the outliers selected during the coincidence step undergo a series of vetoes.
The set of vetoes is applied on all the coincident outliers discarding those having at least one of the following features: they overlap in frequency with known spectral lines (in at least one of the detectors); they are not consistent with the expected significance in each detector (higher significance is expected to arise from the most sensitive detector). These vetoes are described in more detail in the following sections. 

\subsection{Lines Veto and consistency check}
\label{ssec:linesveto}
Due to the presence of instrumental artifacts \cite{DavisdetcharO3}, typically spectral lines \cite{CovaslinesO1O2}), which affect the data quality, outliers lying in a frequency band polluted by a known noise line~\cite{linelist01,virgo3lines} are vetoed. A candidate is then vetoed if during the run its frequency intersects with the frequency region affected by the spectral line. This veto step is applied before coincidences. 
At this stage, $\sim 8.0\%$ of the outliers are removed from the H dataset, while $\sim 4.6\%$ and $\sim 4.9 \%$ are removed from L and V, respectively. 

The list of instrumental artifacts for which the source is not completely understood (e.g. unknown lines) are not used to veto candidates at this stage of the search. 

For the consistency veto \cite{DirectedBSD,allskyO3CW} we have discarded all coincident candidates that show a weighted CR in the less sensitive detector more than three times higher than the one in the more sensitive detector. At this stage the CR we are considering is the one computed from the FH number count. At a later stage, a second consistency veto will be applied using the 5-vector $\mathcal{S}$-statistic \cite{Astone_2010,PhysRevD.89.062008} (see Sec. \ref{ssec:5vecfu}). In practice, assuming that the noise spectral density in the first detector is worse, i.e. $S_{n_1}(f) > S_{n_2}(f)$ , only the outliers with ${\rho_{\rm CR_{1}}}/{\sqrt{S_{n_1}(f)}} < 3 {\rho_{\rm CR_{2}}}/{\sqrt{S_{n_2}(f)}}$ survive to the next post-processing step. In this search, however, this veto only had a very minor effect, removing only one candidate in the HL pair when applied after the lines veto.

At this stage about $37000$ outliers survived for the double coincidences Livingston--Hanford (LH), Livingston--Virgo (LV), Hanford--Virgo (HV), while of the order of $2000$ outliers did pass this first selection for the triple coincidence. Exact numbers are reported in Table \ref{tab:table1}. 
As discussed at the end of Sec. \ref{ssec:method}, we further reduce the number of candidates to pass to the next veto/followup stages according to their significance $\rho_{\rm CR}$ and/or if their frequencies are below 100 Hz (see Table \ref{tab:table1}).

\begin{table}
\caption{
Number of surviving candidates at each stage of the veto chain. Double (HL,LV,HV) and triple (HLV) coincidences are done among candidates surviving the known lines removal. For the HL pairs all the candidates above the CR threshold are followed up, while for the LV and HV pair we do not follow up candidates above 100 Hz. For the HLV case there were only two candidates with $\rho_{\rm CR}(n) > \rho_{\rm CR,thr}$, hence we also follow up the only candidate present below 100 Hz even if its significance is below the threshold. A total of 361 candidates passed to the next step of the analysis described in Sec. \ref{ssec:5vecfu}.}
\begin{ruledtabular}
\begin{tabular}{lccc}
\textrm{} &
\textrm{Single} &
\multicolumn{2}{c}{\textrm{After lines removal}}\\
\colrule
H & 3409029 & \multicolumn{2}{c}{3135640}  \\
L & 3410398 & \multicolumn{2}{c}{3250804}  \\
V & 3408485 & \multicolumn{2}{c}{3240966}  \\
\colrule
\textrm{} & \textrm{Double} & $f_0<100~\rm Hz$ & \textrm{$\rho_{\rm CR}(n) > \rho_{\rm CR,thr}$} \\
HL & 37570 & selection not applied &\bf{274}\\
LV & 37455 & 458 & \bf{40}\\
HV & 36988 & 381 & \bf{44}\\
\colrule
\textrm{} & \textrm{Triple} & &\\
HLV & 2142 & \bf{1} & \bf{2} 
\end{tabular}
\end{ruledtabular}
\label{tab:table1}%
\end{table}

\subsection{Semicoherent 5-vector followup}
\label{ssec:5vecfu}
On the double and triple coincident outliers passing the previous vetoes, we have applied a semi-coherent followup method, based on the so-called \textit{5-vectors}, already used in \cite{SNRO3}, which will be briefly summarized here. 
The method is based on the expected increase of the significance of a given outlier after the frequency modulations are removed from the data, e.g. by applying the heterodyne correction. For searches of CW signals from known pulsars, after removing the effects that modulate the frequency, an amplitude modulation --- due to the sidereal pattern --- still remains. The sidereal modulation pattern, that arises when the signal is integrated for a chunk of data at least equal to the sidereal day, is the key ingredient used to distinguish between an astrophysical signal or a noise outlier.  This modulation produces a splitting of the intrinsic signal angular frequency $\omega_0$ into five frequencies $\omega_0, \omega_0\pm \Omega_{\rm{sid}}, \omega_0 \pm 2\Omega_{\rm{sid}}$, where $\Omega_{\rm{sid}}$ is the Earth’s sidereal angular frequency.

The 5-vector template is then built assuming as known the frequency, spin-down and sky position of the source. In our case we apply a matched filter using the 5-vector shape to chunks of data with duration $T_{\rm{sid}}=2\pi/\Omega_{\rm{sid}}$ equal to one sidereal day, and a final detection statistic $\mathcal{S}$ is computed summing all the detection statistics computed in each chunk. The same calculation is done in the off-source region, i.e. away from the frequency of the candidate. Details on the $\mathcal{S}$-statistic can be found in \cite{SNRO3}. From the $\mathcal{S}$-statistic it is possible to compute the corresponding CR and signal-to-noise ratio (SNR) values. 

We compute the significance of a given outlier using the semi-coherent method described above in two different situations: i) when we remove the Doppler and spin-down modulations (in this case, if the outlier is of astrophysical origin we expect to have a higher significance), and  ii) when no demodulation is applied to the time series. We expect to have an increase of the significance for the case of the demodulated signal when compared to the case where no demodulation is applied. We use these criteria to keep interesting outliers and pass them to the final followup stage. 
After this stage the number of remaining candidates decrease to 62 for the LH pair, 13 and 10 for the HV, LV pairs with frequencies below 100 Hz, respectively. Two out of the three  HLV candidates investigated in this stage have been discarded, while the candidate below 100 Hz passed to the next step.

\subsection{Cumulative significance check}
For this followup stage we rely on the consistency of the significance of a CW signal during the observing time. Specifically, we expect that the significance of a signal will steadily increase as we integrate over more time and hence accumulate more power, following a positive trend as more data is used.
On the other hand, a sudden increase or decrease in the significance as more data is integrated is a clue indicating the presence of non-stationary noise.
We compute the cumulative signal-to-noise ratio and CR on a monthly basis using the 5-vector statistic over a starting time series (no Doppler or spin-down phase correction is applied), and we compare this trend with the one using the heterodyne-corrected time series. 
The heterodyne phase correction applied in this latter case, $e^{i\phi(f_{0},\dot{f_{0}})}$, is fully described by the parameters of the outlier investigated. These curves are computed separately for each detector.
Many outliers presented inconsistencies between the corrected vs the uncorrected case in the same detector (e.g. the cumulative curve of the uncorrected case was always above the corresponding corrected one) or inconsistencies between the two detectors (e.g. for the corrected time series, the cumulative curve of the less sensitive detector was always above the corresponding curve of the most sensitive one). 
The inconsistencies between the curves suggest that these outliers have been produced by an artifact in only one detector. Indeed for the HV and LV pairs, and the HLV triplets, all of the outliers have been discarded with this veto, while 15 candidates for the HL pair were further investigated through visual inspection. We describe these additional tests in the following. 

\subsection{Spectra and peakmaps inspection}
To  further investigate  the remaining 15 outliers from HL, we visually inspect the spectra and the peakmaps using different frequency resolutions. In this way eventually hidden noise features can arise, e.g. narrower spectral lines, and the true origin of the signal candidate can be found. As a second check we have verified if some of the remaining outliers had an evolving frequency that crossed one of the lines of the unidentified set of spectral noise artifacts found in the detectors~\cite{lineunidlist01}, although none of the candidates overlapped with this set of unidentified lines. 
Concerning the spectra we have visually inspected differences and similarities between i) the spectra of the heterodyned time series, corrected using the candidate's parameters $(f_{0},\dot{f_{0}})$, ii) the original spectra, when no correction is applied. If instrumental lines are present these should be present also before correcting the time series. We use three different frequency resolutions for these spectra of $3.2\times 10^{-8}~\rm Hz$, $3.2\times 10^{-7}~\rm Hz$ and $3.2\times 10^{-6}~\rm Hz$, equivalent to full resolution spectra (12 months, full run, no average) 
and to averaging over chunks of duration $T_{\rm obs}/10$, $T_{\rm obs}/100$ respectively.
Concerning the peakmaps, we have checked if local maxima appear in the frequency histogram of the peakmaps (i.e. the peakmap projected onto the frequency axis) in the corrected and/or uncorrected case. Also in this case, if an instrumental line is present, this should appear in the peakmap and peakmap histogram before correction, polluting the interested band. These peakmaps have a frequency resolution of $1/T_{\rm coh}$ while the histograms are built using five different bin widths from $1/T_{\rm coh}$ to $5/T_{\rm coh}$. We use wider bins to be robust against signals which may deviate from the model (or if the candidate's parameters are not accurate enough).
Through this inspection we have been able to discard 13 out of these 15 candidates, given the presence of noise spectral artefacts likely consistent with weak instrumental lines in the frequency band intersected by the candidate, which clearly appear either in the uncorrected spectra or in the peakmap histograms.  

Two out of these 15 visually inspected candidates did not show a clear presence of a weak line nearby before correction, although neither a strong feature suggesting their astrophysical origin after the correction (e.g. a typical 5-vector shape in the corrected spectra). We quantify the significance associated to these candidates using the peakmap histogram counts over the frequency. We consider the frequency subband originally used to select the candidates in the FrequencyHough map. We divide this 0.01 Hz band into subbands of $5$ bins each.
Over each of these smaller frequency subbands we pick the maximum count of the peakmap projection. We then compare the position of the maximum peakmap histogram count of the candidate subband with all the maxima computed on the remaining subbands.

We tag as interesting the candidates ranking first or second among the subbands in both detectors. One of the two surviving candidates ranked 59th in H and 30th in L, hence we discard it. 

The second candidate ranked 11th in H and 1st in L, thus confirming its low significance in the Hanford detector. We ran an additional complementary multi-stage follow-up using the method described in \cite{Tenorio:2021njf} with the PyFstat package \cite{Ashton:2018ure,Keitel:2021xeq} and with the same configuration as in \cite{allskyO3CW}. The resulting Bayes factor was significantly lower than what would be expected for a signal within the probed sensitivity range. The original parameters of the candidates are reported in Table ~\ref{tab:table2}.
Also, the initial threshold used for candidate selection was extremely low, opening the possibility to select such outliers compatible with noise fluctuations. As there is no strong evidence for the presence of an astrophysical signal, we can compute upper limits on the signal strain and from them derive some astrophysical constraints.

\begin{table}
\caption{Parameters of the surviving outlier. The sky position is the one used for the search and coincident to that of Sgr A$^{*}$ with $\alpha_{\rm GC}$ = 17h 45m 40.04s and $\delta_{\rm GC} = -29^{\circ} \,00'\, 28.1"$. The reference time for the parameters is $1238112018$~GPS.}
\begin{tabular}{lccc}
\hline
\textrm{Detector} & $f_0$ [Hz] & $\dot{f_0}$ [Hz/s]& $\rho_{\rm CR}(n)$\\
\hline
H & $908.7708738$ & $-2.511 \times 10^{-9}$ & 5.17\\
L & $908.7704061$ & $-2.521 \times 10^{-9}$ & 4.99\\
\hline
\end{tabular}
\label{tab:table2}%
\end{table}

We conclude that no convincing candidates for a CW signal remain. Hence, we can compute upper limits on the signal strain and from them derive some astrophysical constraints.

\section{Results}
\label{sec:result}
In this section we present the estimates of the upper limits on the signal strain and the constraints we can place in the absence of a detection.
We use a quick method to estimate the upper limits, i.e. the maximum $h_{0}$ allowed by this search, above which we can exclude the presence of a CW signal with a given confidence level.
These limits can be translated into some astrophysical constraints on the ellipticity of neutron stars and the r-mode amplitude. Furthermore, we present for the first time --- for a directed search toward the GC --- ``exclusion regions'' for the boson mass and black hole mass for boson clouds forming around spinning black holes.

\subsection{Upper limits}

We provide an estimate of the upper limits using a method based on the sensitivity estimates presented in \cite{FrequencyHoughmethod}. The minimum detectable strain amplitude $h_{0,\min}$, with $\Gamma=0.95$ corresponding to a $95\%$ confidence level, can be written as:
\begin{equation}
\label{eqn:h0min}
h_{0, \min } \approx \frac{\mathcal{B}}{N^{1 / 4}} \sqrt{\frac{S_{n}(f)}{T_{\rm coh}}}\sqrt{ \rho_{\rm CR,thr}-\sqrt{2} \operatorname{erfc}^{-1}(2 \Gamma)}
\end{equation}
where $N \sim T_{{\rm obs}}/T_{{\rm coh}}$ is the effective number of FFTs used for the search.
In this equation, $\mathcal{B}$ is a parameter that depends on the threshold used for peak selection in the peakmap and on a factor dependent on the sky position of the source and on the signal polarization, which is averaged out. For this search we have computed the value of $\mathcal{B}$ for the GC case. More details are provided in Appendix \ref{sec:appen}. The formula in Eq. \ref{eqn:h0min} expresses the minimum detectable strain by a search which selects the candidate with a CR higher than $\rho_{\rm CR,thr}$, i.e. it states the best sensitivity a given search can achieve when the minimum CR of our candidates coincides with $\rho_{\rm CR,thr}$. On the other hand, if we want to look for the maximum allowed $h_0$, above which we can exclude the presence of a CW signal, i.e. provide an estimate of the upper limit, we can substitute for $\rho_{\rm CR,thr}$ the value of the maximum CR $(\rho_{\rm CR,max})$ found in a given frequency band. The width of the 1-Hz frequency band determines the resolution of the upper limits estimate. 

We have verified that this estimate of the upper limits, using $\rho_{\rm CR,max}$ in Eq. \ref{eqn:h0min}, already implemented in \cite{allskyO3CW,allskyO3BC}, indeed yields conservative limits if compared to the results obtained with the classical frequentist approach using artificially injected signals. This check has been performed on six frequency bands of 1 Hz each,  randomly chosen over the full frequency range investigated. We have also verified that the 95\% confidence level upper limits, obtained using software injected signals, are always above the curve defined in Eq. \ref{eqn:h0min} when $\rho_{\rm CR,thr}$ is used. 
Furthermore the difference between the upper limits obtained using injections and those computed using this method is still within the calibration uncertainty errors, which are already affecting the estimate of the noise power spectral density $S_n(f)$ (see Sec. \ref{sec:data}~\cite{SunO3acalibration2020,SunO3bcalibration2021,AcerneseO3calibration2021}). 

\begin{figure*}
\includegraphics[scale=0.55]{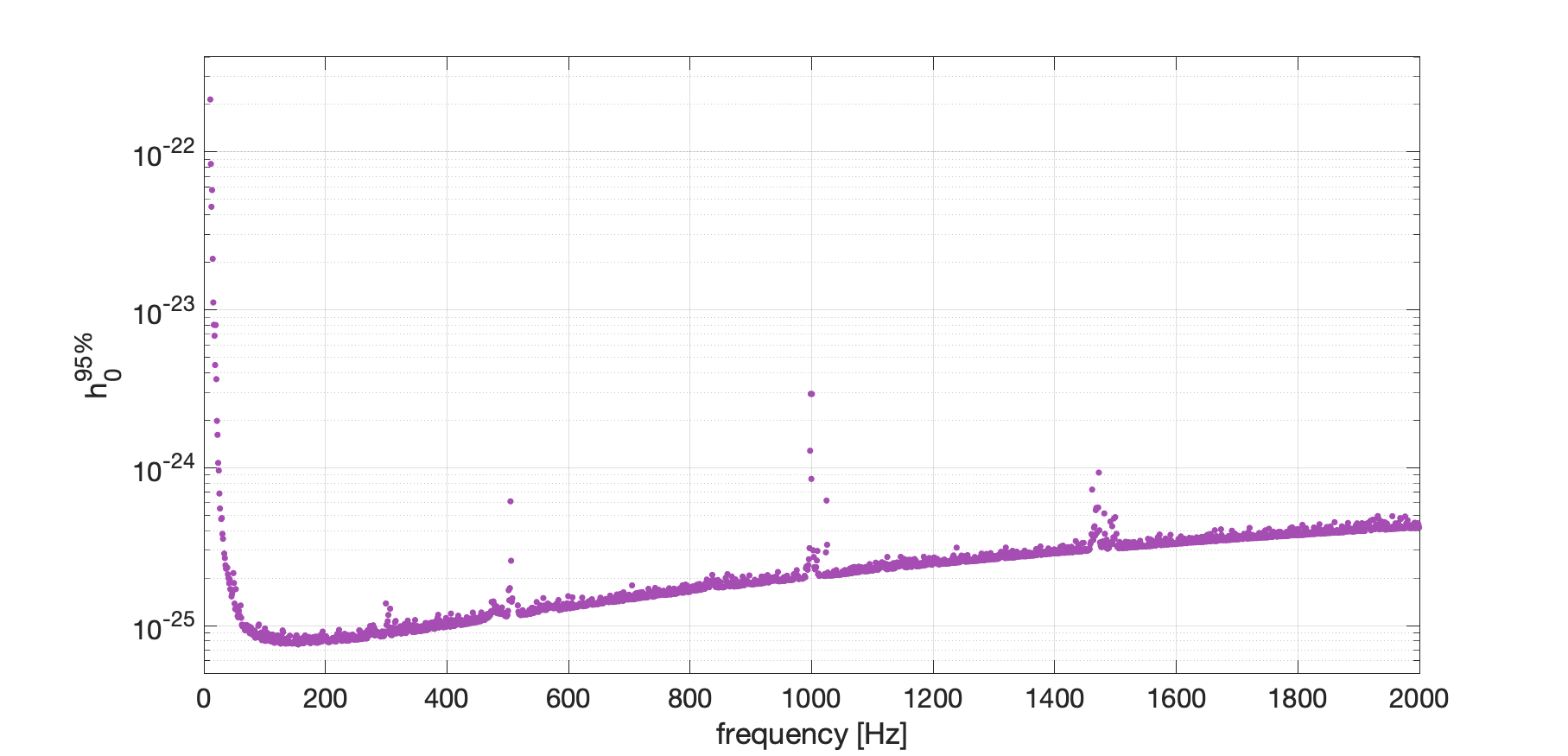}
\caption{Estimates of the 95 $\%$ C.L. strain upper limits, derived for the best combination HL in 1 Hz bands.}
\label{fig:upperlimit} 
\end{figure*}

The curve shown in Fig. \ref{fig:upperlimit} represents the joint upper limits estimate of the coincident pair HL. This is obtained by computing the $h_{0}^{95\%}(\rho_{\rm CR,max},S_n)$ separately for each H and L detector and keeping the worst of the two curves in each Hz. For a given detector, the CR has been taken equal to the maximum in each 1-Hz band, otherwise was set equal to $\rho_{\rm CR,thr}$ if the maximum CR in that band was lower than $\rho_{\rm CR,thr}$. 
Given that for a given pair (or triplet) of detectors the combined upper limits curves are dominated by the less sensitive detector, we do not report the LV, HV, LHV associated curves given the large difference between the $S_n(f)$ of the Virgo detector compared to the power spectral density of Hanford and Livingston. 
From the curve in Fig. \ref{fig:upperlimit} it is possible to see a minimum strain of $\sim 7.6 \times 10^{-26}$ at 140 Hz. This result improves on the one in \cite{DirectedBSD} using O2 data by a factor $\simeq 1.9$.

\subsection{Astrophysical implications}
We can exploit the relation between the GW strain amplitude and some astrophysical parameters characterizing the emitting system to derive some constraints. In particular, for the isolated neutron star case, we can map the $h_0^{95\%}$ upper limit curve to a constraint on the ellipticity of the the star \cite{Johnson-McDaniel2013}, making some assumptions on the moment of inertia of the spinning star. It is also possible to convert the upper limits on the strain into constraints on the  r-mode amplitude as discussed in \cite{Owen2010}, assuming a coherent emission during the observing time. In addition to these classic limits, typically derived for CW searches from neutron stars, we can derive some exclusion regions over the masses involved 
in a superradiance process of boson particles around spinning black holes as discussed in Sec. \ref{sssec:bosons}.

\subsubsection{Neutron stars}

\paragraph{Ellipticity}
For the prototypical case of a rotating neutron star with non-axisymmetric deformations misaligned with the rotation axis, the strain amplitude is proportional to twice its spin frequency, $f_{\rm GW}=2f_{\rm spin}$. This scenario corresponds e.g. to the presence of mountains on the neutron star's surface~\cite{Gittins2020} and the strain amplitude can be written as
\begin{equation}
\label{eq:h0}
h_{0}=\frac{4 \pi^{2} G}{c^{4}} \frac{I_{z z} f_{\rm GW}^{2}}{d} \epsilon.
\end{equation}

Assuming that the moment of inertia for a perpendicular biaxial rotor spinning around z, $I_{zz}=qI_{\rm fid}$, can be a multiple of the fiducial moment of inertia $I_{\rm fid}=10^{38}$~kg\,m$^2$, where $q$ is a proportionality factor\footnote{The exact value of a neutron star's moment of inertia is unknown and it strongly depends on its EOS, for this reason we will use a proportionality factor $q=1$ (normal matter)  and $q=5$ (extreme matter) to cover these two extremes, see \cite{Johnson-McDaniel2013}},  the $h_0$ can be used to express the ellipticity of the neutron star as a function of the GW signal frequency and the moment of inertia as 

\begin{equation} 
\label{eqn:eps} 
\epsilon = 7\times 10^{-4}\left(\frac{I_{\rm fid}}{I_{zz}}\right)\left(\frac{h_0}{10^{-24}}\right)\left(\frac{100{\rm~Hz}}{f_{\rm GW}}\right)^2.
\end{equation}

\begin{figure}
\includegraphics[scale=0.46]{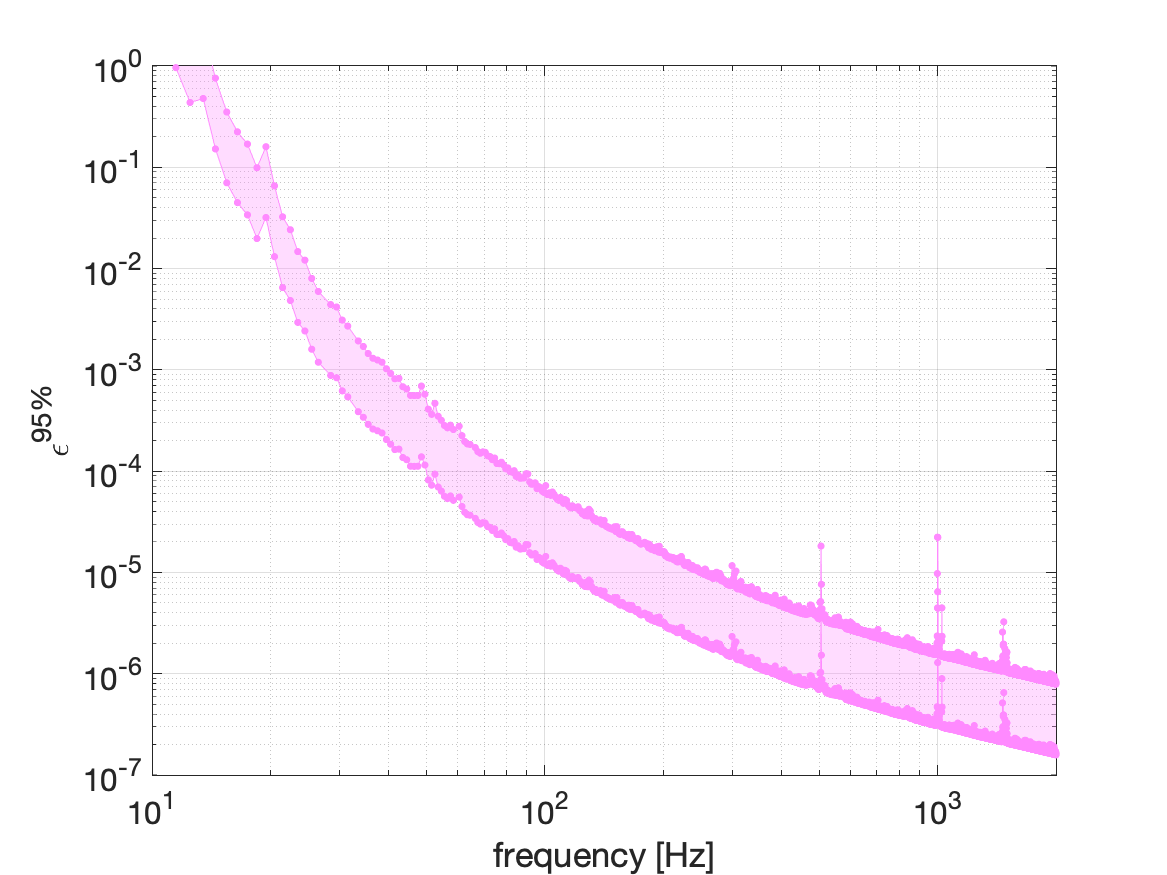}
\caption{Estimates of the 95 $\%$ C.L. ellipticity upper limits assuming a GC distance of 8 kpc. The shaded area between the two curves covers the possible values of the moment of inertia along  z of the spinning star. The lower curve corresponds to a moment of inertia five times larger than the fiducial value $I_{\rm fid}$, sustainable by exotic objects.}
\label{fig:epsilon}
\end{figure}

Here a distance of the GC of $d=8~\rm{kpc}$ has been assumed, although different estimates of $d$ exist~\cite{Abuter2019,Eckart2017,distanceGC,VERAGCdist2020}. 
In Fig. \ref{fig:epsilon} we report the estimated 95\% confidence level upper limits of the ellipticity, $\epsilon^{95\%}$. The two curves indicate the two extreme cases for $I_{zz}=I_{\rm fid}$ (upper curve), and $I_{zz}=5I_{\rm fid}$ (lower curve). The shaded region spans all the possibilities between the $q=1$ and $q=5$ case. Moments of inertia five times larger than the fiducial value can possibly be sustained by stars made up of more exotic components.  
For the $q=1$ case the minimum ellipticity reached for the highest frequency is $\epsilon=7.26\times 10^{-7}$, while a minimum of $\epsilon=1.45 \times 10^{-7}$ is obtained for the $q=5$ case.
Given this high uncertainty of the actual moment of inertia of the star (dependent on both the mass and the radius of the star), it is useful to quote the corresponding mass quadrupole $Q_{22}$ component of the $l = m = 2$ mode, which is present in the expression of the GW amplitudes in the mass quadrupole formalism \cite{Owen2005}

\begin{equation}
\label{eqn:Q22} 
Q_{22} = \sqrt{\frac{15}{8\pi}}\epsilon I_{zz}.
\end{equation}

This quantity is then independent of the actual moment of inertia used and is directly connected with $h_{0}$ (see Eq. \ref{eq:h0}).
A minimum value of $Q_{22}=5.61 \times 10^{31}~\rm{kg~m^2}$ is reached at the highest frequency. 

In this simplest model we are not considering the possible multiple harmonic emission mechanism active in situations like a superfluid pinned to the crust, a triaxial star not spinning around its principal axis and more \cite{Jones2010,Jones2015}, where additional radiation is expected at the star's spin frequency $f_{\rm spin}$ as well at the $2f_{\rm spin}$ frequency. The case of free precession would require including
further terms in addition to the dual harmonic components \cite{Jones2001}. \\

\paragraph{R-mode amplitude}
Changing the emission scenario but still staying in the single harmonic emission model, the limits on the strain can be parametrized for the case of unstable oscillation modes, namely r-modes, happening at $f_{\rm GW}=\frac{4}{3}f_{\rm spin}$ in the non-relativistic case. The actual proportionality factor between $f_{\rm spin}$ and $f_{\rm GW}$ is expected to differ from $\frac{4}{3}$ in the relativistic case and when the EOS dependence is considered (see \cite{Andersson2014,Idrisy2015}).

Following the discussion in \cite{Owen2010} it is possible to convert upper limits from the ``mountain" scenario to the equivalent r-mode expression, given that the latter can be obtained from the standard expression in the ellipticity case through the mapping $\psi \rightarrow \psi+\pi / 4$ corresponding to a 45 degree rotation of the polarization angle $\psi$. The amplitude of r-mode emissions is then given by
\begin{equation}
\label{eqn:alpha}
\alpha \simeq 0.028\left(\frac{h_{0}}{10^{-24}}\right)\left(\frac{d}{1~\rm kpc}\right)\left(\frac{100 \rm{~Hz}}{f_{\rm GW}}\right)^{3}.
\end{equation}

\begin{figure}
\includegraphics[scale=0.46]{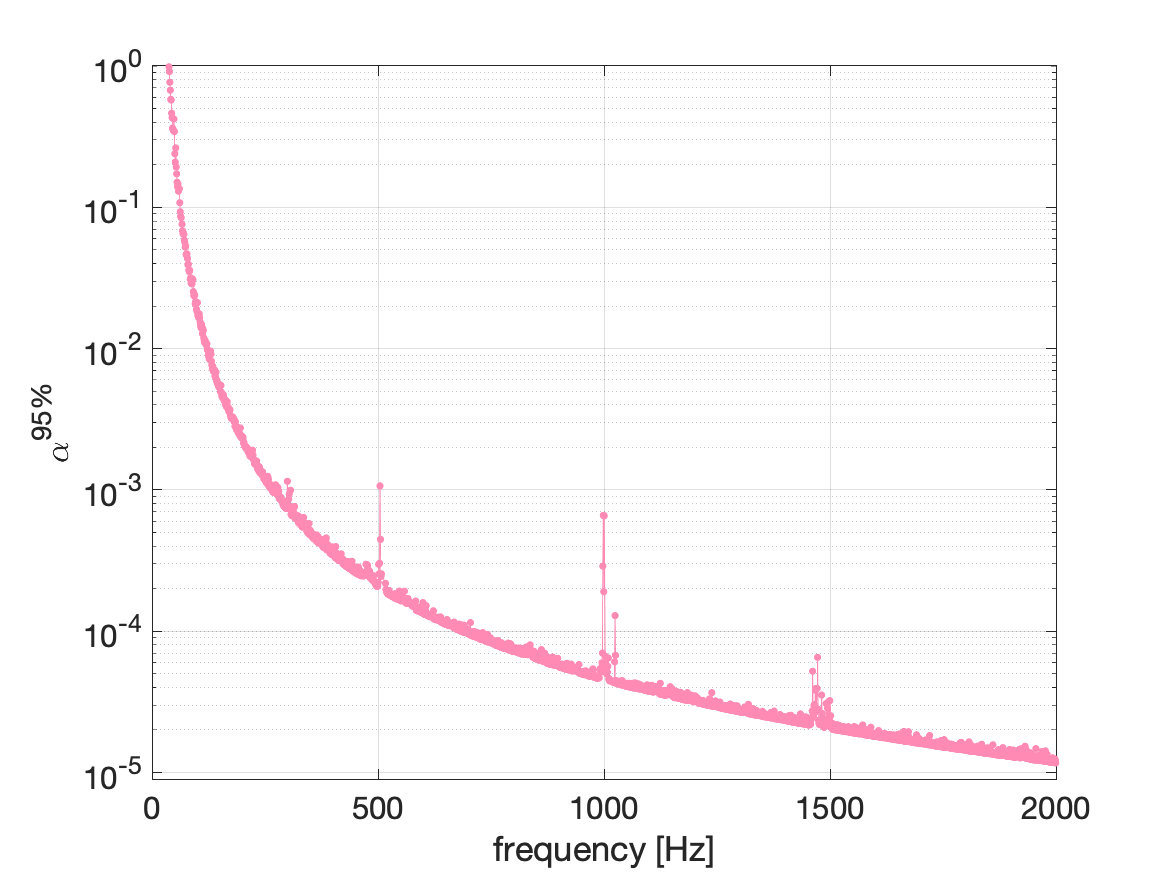}
\caption{Estimates of the 95 $\%$ C.L. r-mode amplitude upper limits for neutron stars in the GC region assuming $d=8$ kpc.}
\label{fig:alpha}
\end{figure}

This equation has been obtained assuming the dimensionless functional of the neutron star EOS $\tilde{J} \approx 0.0164$ and a neutron star with mass $M=1.4 M_{\odot}$ and radius $R=11.7~\rm km$ as in \cite{Owen2010}. We report the converted $95\%$ C.L. upper limits on the r-mode  amplitude $\alpha$ in Fig. \ref{fig:alpha}, assuming a GC distance of 8 kpc as done for Fig. \ref{fig:epsilon}.  Minimum values of $\alpha \sim 10^{-5}$ are reached for the highest frequencies.

\subsubsection{Boson clouds}
\label{sssec:bosons}
A population of thousands of stellar-mass black holes is expected to exist near the GC \cite{Freitag2006} and observational evidence is accumulating, see e.g.  \cite{Hailey2018,Mori2021}. A fraction of these black holes may have developed a boson cloud \cite{Brito:2017zvb,Arvanitaki2015,Baryakhtar2021,BritoSuperradiance2020} which is currently depleting by emitting CWs. 

In order to put constraints on the boson cloud systems we have first derived the upper limits curve in Fig. \ref{fig:upperlimit}, using only candidates with a positive spin-down (e.g. with spin-up). This is a necessary step to have a more accurate estimate to use for our constraints, since the signal produced by boson clouds around spinning black holes is characterized by a spin-up \cite{Arvanitaki2015} and no spin-down is expected as for the case of spinning neutron stars. 

\begin{figure}
\includegraphics[scale=0.30]{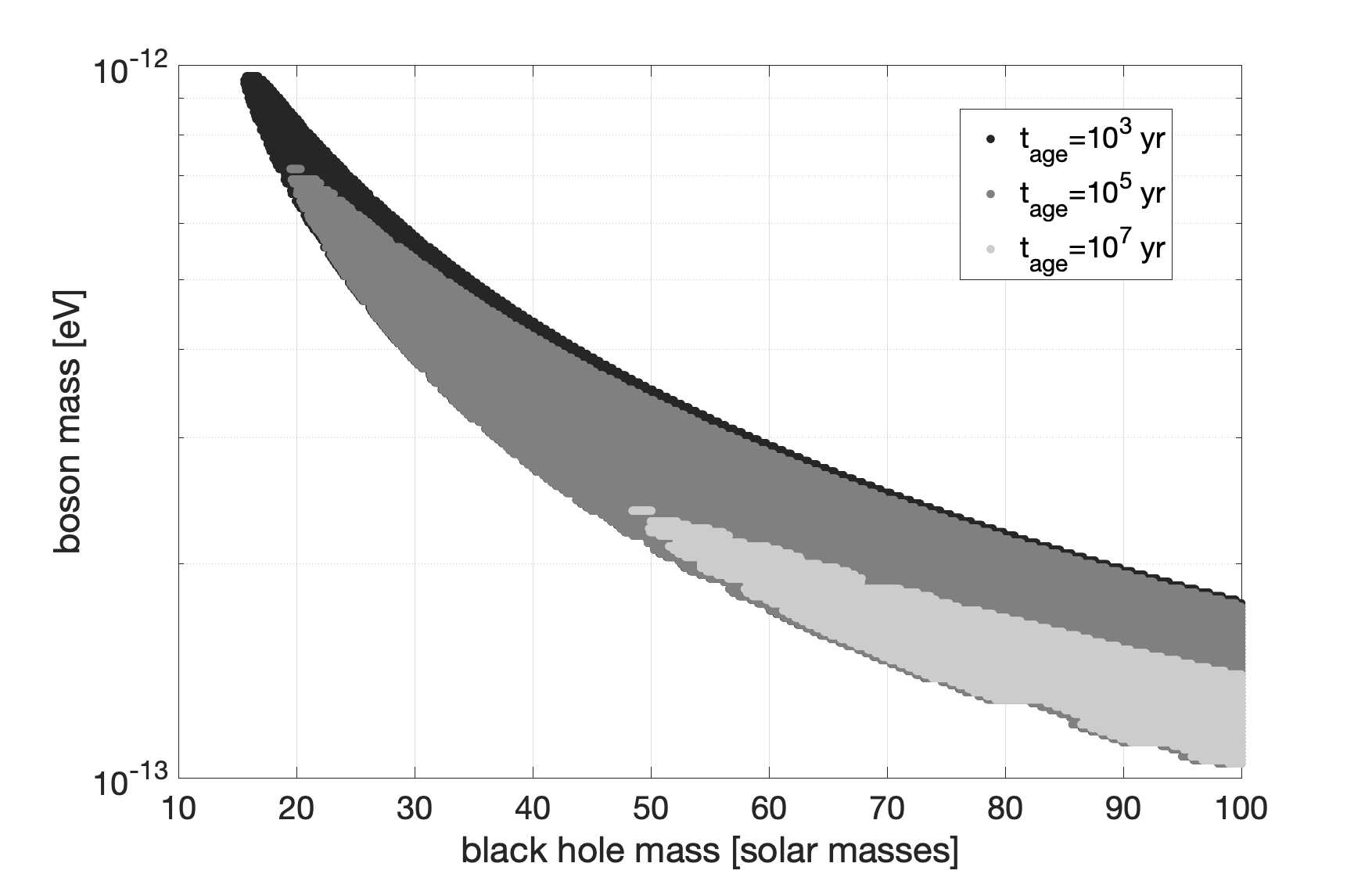}
\caption{Constraints on the black hole mass -- boson mass plane, assuming CW emission from boson clouds around spinning black holes located in the GC. An initial black hole dimensionless spin $\chi_i=0.5$ and cloud ages of $t_{\rm age}=10^3,~10^5,~10^7$ years are considered.}
\label{fig:boson_constr} 
\end{figure}

Given an upper limits curve, we can translate it into constraints in the black hole/boson mass parameter space. Following what was done in previous all-sky searches \cite{BosonsPRL2019,allskyO3BC}, we compute ``exclusion regions'' in the parameter space for a given value of the black hole spin before the superradiant cloud growth $\chi_i$ and for different boson cloud ages, $t_{\rm age}$. The distance is fixed to 8 kpc. Specifically, Fig. \ref{fig:boson_constr} shows the constraints for 
$\chi_i=0.5$ and $t_{\rm age}=10^3,~10^5,~10^7$ years.
Going from these constraints, valid for specific parameter choices, to the actual exclusion of given combinations of black hole and boson masses is not trivial, as it depends on the uncertain characteristics of the black hole population in the GC. In particular, it is expected that many black holes now residing in the GC region may have ages of Gyrs \cite{Emami2020ObservationalSO}, and then would not be relevant anymore from the CW emission point of view. On the other hand, a non-negligible number of black holes should have formed more recently, both by core collapse of a progenitor massive star, or by the coalescence of black hole binary systems, formed e.g. by tidal capture in the dense GC environment \cite{2018MNRAS.478.4030G}. These systems could have developed a boson cloud which is still in the CW emission phase. A quantitative study of this subject is clearly important but outside the scope of the current paper.   

\section{Conclusion}
\label{sec:conc}
We have presented a search for continuous GWs from sources in the GC using LIGO and Virgo data from the third observing run. 
Although the core of this search is the same as the one presented in  \cite{DirectedBSD}, a more sensitive, longer data set has been used, and several novelties have been introduced in this version.
First of all the parameter space investigated is much wider, allowing for high-frequency emitters to be searched for.  In addition, data from the Virgo detector has been used for the first time, providing an increased number of potential candidates. 
Along with these extensions, new techniques have been applied in the followup part and for the computation of upper limits. The enormous reduction of the computational load of this last step of the analysis allows for a better use of the resources in the followup part. Indeed this makes possible the use of a lower threshold in the first pass selection of candidates,  increasing the final number of outliers, the chance of detection and the overall search sensitivity. One marginal outlier near  909~Hz (see Table \ref{tab:table2}) could not be decisively ruled out, but is more consistent with a single-detector noise fluctuation.

The deepest strain limit is $7.6 \times 10^{-26}$ at 142 Hz, corresponding  to levels of ellipticity well below the maximum value expected for a neutron star composed of standard matter 
\cite{Ushomirsky2002,Haskell2006,Johnson-McDaniel2013}, solid strange stars, or hybrid and meson-condensate stars \cite{Owen2005} for most of the frequency band investigated. The most stringent constraints on the r-mode amplitude are obtained at the highest frequencies well below  expected quantities for the nonlinear saturation mechanisms \cite{Bondarescu2009}.
Finally we provide new constraints on the mass distribution of boson cloud masses in addition to those presented in \cite{allskyO3BC}.

\appendix

\section{Details on upper limits formula}
\label{sec:appen}
Following the computation of equation 67 in \cite{FrequencyHoughmethod} we want to compute the average prefactor $\mathcal{B}$ in Eq. \ref{eqn:h0min} of this paper. The corresponding expression for  $\mathcal{B}$ for the all-sky case described in \cite{FrequencyHoughmethod} is equal to

\begin{equation}
\mathcal{B}_{\rm all-sky}= \frac{4.02}{\theta_{\rm thr}^{1/2}} \left(\frac{p_{0}\left(1-p_{0}\right)}{p_{1}^{2}}\right)^{1/4},
\end{equation}
where the number $4.02$ results from the average of all the varying quantities $\alpha$, $\delta$, $\psi$ (see Eq. B19 in \cite{FrequencyHoughmethod}). All the remaining quantities, i.e. the peak selection threshold $\theta_{\rm thr}=2.5$,  the probability to select a noise peak $p_0=0.0755$, as well $p_{1}=0.0692$, a function depending on $\theta_{\rm thr}$, remain unchanged from \cite{FrequencyHoughmethod} in this search. 
For our problem we need to evaluate the expression for a particular sky position and no averaging over $\alpha$ and $\delta$ is needed. On the other hand,  we do need to average over time and the polarization parameter $\psi$, which is uniformly distributed over $[-\pi/4,\pi/4]$.
Let us now consider the expressions for the beam pattern functions $F_{+}$ and $F_{\times}$ \cite{Jaranowski1998}

\begin{equation}
\begin{aligned}
F_{+}(\psi,t)=a(t) \cos{2\psi}+b(t) \sin{2\psi}, \\
F_{\times}(\psi,t)=b(t) \cos{2\psi}  - a(t) \sin{2\psi}.
\end{aligned}
\end{equation}

It is easy to check that the squared average over the polarization angle $\psi$ is equal to 0.5. Hence we can write the squared average of $F_{+}$ and $F_{\times}$ as

\begin{equation}
\label{eqn:beampat}
\left \langle F_{+}^{2}\right\rangle_{\psi,t}=\left\langle F_{+}^{2}\right\rangle_{\psi,t}=\frac{1}{2}\left(\left \langle a^2 \right\rangle_{t} +\left \langle b^2\right\rangle_{t}\right).
\end{equation}

Removing also the dependency over time and considering the following average expressions for $a^2(t)$ and $b^2(t)$ evaluated for the case $T_{\rm obs}=n 2 \pi / \Omega_{\rm sid}$, i.e. an integer number of sidereal days \cite{astone2002allsky}:

\begin{equation}
\begin{aligned}
\left\langle a^{2}\right\rangle_{t}=& \frac{1}{16} \sin ^{2} 2 \gamma\left[9 \cos ^{4} \phi \cos ^{4} \delta+\frac{1}{2} \sin ^{2} 2 \phi\right.\\
&\left.\times \sin ^{2} 2 \delta+\frac{1}{32}(3-\cos 2 \phi)^{2}(3-\cos 2 \delta)^{2}\right] \\
&+\frac{1}{32} \cos ^{2} 2 \gamma\left[4 \cos ^{2} \phi \sin ^{2} 2 \delta\right.\\
&\left.+\sin ^{2} \phi(3-\cos 2 \delta)^{2}\right], \\
\left\langle b^{2}\right\rangle_{t}=& \frac{1}{32} \sin ^{2} 2 \gamma\left[(3-\cos 2 \phi)^{2} \sin ^{2} \delta\right.\\
&\left.+4 \sin ^{2} 2 \phi \cos ^{2} \delta\right]+\frac{1}{4} \cos ^{2} 2 \gamma \\
& \times(1+\cos 2 \phi \cos 2 \delta) .
\end{aligned}
\end{equation}

Evaluating them for the specific sky location $\delta=\delta_{\rm GC}$ and for each $(\gamma,\phi)$ of a given detector, where $\phi$ is the latitude of the detector's site and $\gamma$ is the orientation of the detector's arms, we can obtain different values for the antenna pattern functions per detector ($i-$det):
\begin{equation}
\begin{aligned}
\left.  \left \langle F_{+}^{2} \right \rangle_{\psi,t} \right|_{\delta_{\rm GC},i-{\rm det}}=\left.  \left \langle F_{\times}^{2} \right \rangle_{\psi,t} \right|_{\delta_{\rm GC},i-{\rm det}}\\
=\frac{1}{2}\left(\left.\left \langle a^{2} \right\rangle_{t}\right|_{\delta_{\rm GC},i-{\rm det}}+ \left.\left \langle b^2 \right\rangle_{t}\right|_{\delta_{\rm GC},i-{\rm det}}\right).
\end{aligned}
\end{equation}

The exact numbers entering as a pre-factor of Eq. (67) of  \cite{FrequencyHoughmethod} for this search differ from the value used for the all-sky search in \cite{FrequencyHoughmethod}, where an average over the sky position is considered, by no more than the $~3\%$. 
To be more accurate, the actual pre-factors to be used in the calculation of our upper limits will be 4.06, 4.05 and 4.12 for H, L and V, respectively, while for the all-sky case it is equal to 4.02. This means that $\mathcal{B}$ in Eq. \ref{eqn:h0min} is equal to 5.06, 4.93, 5.08 for H, L and V respectively, while it is equal to 4.97 for the all-sky case.     

\section*{Acknowledgments}
This material is based upon work supported by NSF’s LIGO Laboratory which is a major facility
fully funded by the National Science Foundation.
The authors also gratefully acknowledge the support of
the Science and Technology Facilities Council (STFC) of the
United Kingdom, the Max-Planck-Society (MPS), and the State of
Niedersachsen/Germany for support of the construction of Advanced LIGO 
and construction and operation of the GEO\,600 detector. 
Additional support for Advanced LIGO was provided by the Australian Research Council.
The authors gratefully acknowledge the Italian Istituto Nazionale di Fisica Nucleare (INFN),  
the French Centre National de la Recherche Scientifique (CNRS) and
the Netherlands Organization for Scientific Research (NWO), 
for the construction and operation of the Virgo detector
and the creation and support  of the EGO consortium. 
The authors also gratefully acknowledge research support from these agencies as well as by 
the Council of Scientific and Industrial Research of India, 
the Department of Science and Technology, India,
the Science \& Engineering Research Board (SERB), India,
the Ministry of Human Resource Development, India,
the Spanish Agencia Estatal de Investigaci\'on (AEI),
the Spanish Ministerio de Ciencia e Innovaci\'on and Ministerio de Universidades,
the Conselleria de Fons Europeus, Universitat i Cultura and the Direcci\'o General de Pol\'{\i}tica Universitaria i Recerca del Govern de les Illes Balears,
the Conselleria d'Innovaci\'o, Universitats, Ci\`encia i Societat Digital de la Generalitat Valenciana and
the CERCA Programme Generalitat de Catalunya, Spain,
the National Science Centre of Poland and the European Union – European Regional Development Fund; Foundation for Polish Science (FNP),
the Swiss National Science Foundation (SNSF),
the Russian Foundation for Basic Research, 
the Russian Science Foundation,
the European Commission,
the European Social Funds (ESF),
the European Regional Development Funds (ERDF),
the Royal Society, 
the Scottish Funding Council, 
the Scottish Universities Physics Alliance, 
the Hungarian Scientific Research Fund (OTKA),
the French Lyon Institute of Origins (LIO),
the Belgian Fonds de la Recherche Scientifique (FRS-FNRS), 
Actions de Recherche Concertées (ARC) and
Fonds Wetenschappelijk Onderzoek – Vlaanderen (FWO), Belgium,
the Paris \^{I}le-de-France Region, 
the National Research, Development and Innovation Office Hungary (NKFIH), 
the National Research Foundation of Korea,
the Natural Science and Engineering Research Council Canada,
Canadian Foundation for Innovation (CFI),
the Brazilian Ministry of Science, Technology, and Innovations,
the International Center for Theoretical Physics South American Institute for Fundamental Research (ICTP-SAIFR), 
the Research Grants Council of Hong Kong,
the National Natural Science Foundation of China (NSFC),
the Leverhulme Trust, 
the Research Corporation, 
the Ministry of Science and Technology (MOST), Taiwan,
the United States Department of Energy,
and
the Kavli Foundation.
The authors gratefully acknowledge the support of the NSF, STFC, INFN and CNRS for provision of computational resources.

This work was supported by MEXT, JSPS Leading-edge Research Infrastructure Program, JSPS Grant-in-Aid for Specially Promoted Research 26000005, JSPS Grant-in-Aid for Scientific Research on Innovative Areas 2905: JP17H06358, JP17H06361 and JP17H06364, JSPS Core-to-Core Program A. Advanced Research Networks, JSPS Grant-in-Aid for Scientific Research (S) 17H06133 and 20H05639 , JSPS Grant-in-Aid for Transformative Research Areas (A) 20A203: JP20H05854, the joint research program of the Institute for Cosmic Ray Research, University of Tokyo, National Research Foundation (NRF), Computing Infrastructure Project of KISTI-GSDC, Korea Astronomy and Space Science Institute (KASI), and Ministry of Science and ICT (MSIT) in Korea, Academia Sinica (AS), AS Grid Center (ASGC) and the Ministry of Science and Technology (MoST) in Taiwan under grants including AS-CDA-105-M06, Advanced Technology Center (ATC) of NAOJ, and Mechanical Engineering Center of KEK.



\newpage
\bibliographystyle{ieeetr}
\bibliography{references}

\iftoggle{endauthorlist}{
  \let\author\myauthor
  \let\affiliation\myaffiliation
  \let\maketitle\mymaketitle
  \title{The LIGO Scientific Collaboration, Virgo Collaboration, and KAGRA Collaboration}
  \pacs{}
    
  \newpage
  \maketitle
}

\end{document}